\documentclass[12pt]{article}
\usepackage{epsf}
\title{Small size pentaquark in QCD sum rule approach}

\author{A.G.Oganesian\\
Institute of Theoretical and Experimental Physics,\\
B.Cheremushkinskaya 25, 117218 Moscow,Russia}
\date{}
\begin{document}

\maketitle

\newcommand{\be}{\begin{equation}}
\newcommand{\ee}{\end{equation}}

\def\la{\mathrel{\mathpalette\fun <}}
\def\ga{\mathrel{\mathpalette\fun >}}
\def\fun#1#2{\lower3.6pt\vbox{\baselineskip0pt\lineskip.9pt
\ialign{$\mathsurround=0pt#1\hfil##\hfil$\crcr#2\crcr\sim\crcr}}}

\begin{abstract}
 The analysis of the sum rules (for two different
 choices of 5-quark currents) indicate that to
explain experimental results it is necessary to have some
one-particle lower state (and pentaquark is a reliable candidate
to this state) while the two-particle coupled lower state ($NK$
system) should be excluded. Obtained sum rules have good stability
and correspond to pentaquark mass $m_\theta$=1.54 $GeV$ at the
appropriate choice of continuum threshold. Also the features of
the sum rules for 2-point correlator with 5-quark currents are
studied. This features should be taken into account to obtain sum
rules correctly.
\end {abstract}

\newpage
{\bf \large ~1.Introduction}

\vspace{1cm}

Recently, the exotic narrow ( with the width less than 9MeV)
baryon resonance with the quark content $ uudd\bar{s}$ and mass
1.54 GeV $\cite{1}$, $\cite{2}$ had been discovered. It was
supposed that it can be identified with $\Theta^+$ baryon,
predicted in 1997 by D.Diakonov, V.Petrov and M.Polyakov
$\cite{3}$ in the Chiral Soliton Model as a member of
antidecouplet with the hypercharge $Y=2$.

Later, the existence of this resonance was confirmed by many other
groups (see for the review $\cite{4}$). The phase analysis leads to
the conclusion that the width should be even smaller (less than
1MeV, see $\cite{5}$,$\cite{6}$,$\cite{7}$). On the other side, in
a number of experiments searches for this state were unsuccessful.
(see $\cite{4}$ for the review). So one can say, that the question
does pentaquark exist or no has yet no experimental answer.

The explanation of small width of pentaquark was suggested in
papers $\cite{8}$,$\cite{8a}$. It was shown, that if the
pentaquark size is not larger than usual hadron size and its field
is represented by the local quantum field opperator, then its
decay width should be strongly suppressed by chirality
conservation for any choice of the interpolating current. Later
the numerical estimations of pentaquark width were also obtained in QCD
sum rule approach $\cite{9}$,$\cite{9a}$,$\cite{9b}$. In this
paper we will analyze, if is it possible to predict something
about the pentaquark existence using QCD sum rule.

QCD sum rules (S.R.)  is one of the powerful methods of analyzing
such problem, and it is well-known, that  using the sum rules
for two-point correlators it became possible to predict the masses
of usual baryo0ns (see $\cite{10}$). So, reasonably  the question
appears, is it possible to use the same approach for pentaquark
case.

A number of papers are devoted to this question (see, for example
$\cite{11}$-$\cite{16}$). Unfortunately, except $\cite{16}$, the
analysis of the 2-point correlator in the main part of this paper
was restricted at comparatively low dimension operators of
operator product expansion (OPE) series (dimension $d<8$). As was
noted in papers $\cite{8}$,$\cite{16}$, for the case of 5-quark
current the contributions of high dimension operators are
significant and should be taken into account. This is a
consequence of high dimension of 5-quark current. We will discuss
this in sect. 2.

On the other point of view, such significant role of high operator
contribution leads us to a conclusion, that the accuracy of S.R
for multiquark system can not be good (particularly S.R. for
pentaquark mass). Moreover, there is also one source of
uncertainty, which is also absent in the case of usual hadrons. The
problem is, that unlike the usual hadron case, where the masses of
first resonance are considerably smaller than the threshold of
multiparticle states( which are included into continuum) in the
case of pentaquark the threshold of 2-particle intermediate state
($NK$, i.e nucleon and kaon) generation is even a bit smaller than
a pentaquark mass. In this paper we will try to analyze all these
problems for the pentaquark 2-point sum rules.

The following problems will be discussed in this paper:

1. We will discuss  the properties of operator expansion for
5-quark current, and give some arguments, that the main
contribution to S.R. give high dimension operators, but always
there should be a strong numerical cancellation of neighboring
dimensions contributions for any choice of current. It will be
shown (and we will discuss the examples) that contribution of the
dimensions $6$ and $8$ contribution cancel each other up to
$90\%$,  and the same cancellation, though not so strong, but also
significant appears for higher dimension.

2. It will be supposed that $NK$ system contribution in the
spectral density at the region close to the pentaquark mass can be
interpolated by some local current, and it will be checked if it
is possible to saturate the 2-point sum rules only by this system
plus continuum. If yes, then one should say, that 2-point S.R.
results for 5-quark current can not be treated as an argument in
the favor of pentaquark existence, but can be easily explained as
$NK$ system. If no, then we should say, that S.R. required some
one-particle state in the region of masses about $1-2GeV$ (though
the accuracy of determination of mass in any case will be not
good). All the analysis will be done on an example of two
different currents with pentaquark quantum numbers, but we believe
that the obtained conclusion are reliable for any currents.

\vspace{1cm}

{\bf \large ~ 2. General properties of 5-quark current correlators}

\vspace{1cm}

In this section we will discuss some feature of sum rules for 5-quark current correlator.

\be
\Pi(p^2) =i \int e^{ipx} \langle 0 \mid T \eta_{\theta} (x)
\eta_{\theta}(0) \mid 0 \rangle =\hat{p} \cdot \Pi_1(p^2)+\Pi_2
(p^2)\ee

Here $\eta_{\theta}$ is current with quantum numbers of pentaquark and
with quark content $uudd\bar{s}$

Let us define
\be
 \langle 0 \vert \eta_{\Theta} \vert \Theta \rangle = \lambda
 \upsilon_{\Theta}
 \ee

where  $\upsilon_{\Theta}$   is the $\theta^+$   wave function
and $\lambda$ is the corresponding coupling
constant. Then, using narrow resonance approximation and
quark-hadron duality, one can easily write down for physical
representation of invariant amplitudes $\Pi_1(p^2$    and
$\Pi(p^2)$ in (1)

\be
Im \Pi^{phys}_1 =\pi \lambda^2 \delta (p^2-m^2_{\theta})+ \theta
(p^2-s_0) Im \Pi_1^{QCD}\ee

\be
Im \Pi^{phys}_2= \pi\lambda^2\delta (p^2-m^2_{\theta})\cdot
m_{\theta} + \theta(p^2-s_0)Im\Pi_2^{QCD}\ee

where $\Pi^{QCD}$ means the QCD calculation of the corresponding
invariant amplitude, $s_0$ is the continuum threshold and
$m_{\theta}$ is the  $\theta^+$  mass. Then, equating physical
representation and result of QCD calculation one can easily get
(after Borel transformation) two sum rules for $\Pi_1(p^2)$ and
$\Pi(p^2)$ (SR1 and SR2 correspondingly).

\be
\frac{1}{\pi} \int due^{-u/M^2} Im\Pi_1^{QCD}(u) =\lambda^2
e^{-m^2_{\theta}/M^2}\ee

\be
\frac{1}{\pi} \int due^{-u/M^2} Im\Pi_2^{QCD}(u) =\lambda^2
e^{-m^2_{\theta}/M^2} \cdot m_{\theta}\ee

Here  $Im\Pi^{QCD}$  mean the sum of OPE series, i.e.
$$Im\Pi^{QCD}=\sum Im\Pi^{QCD}_d$$ where  $Im\Pi^{QCD}_d $means
the corresponding dimension $d$ contribution in the operator
expansion. (see Fig.1, where examples of diagrams for different
dimension operators are given). We will account terms up to
dimension $d=14$ for SR1 (eq.5) and $d=13$ for SR2 (eq.6). In the
calculation  we will use the quark propagator expansion in OPE up
to $d=7$ operators (the first term of the expansion on $s$-quark
mass will be also accounted)

\be
\langle 0\mid T q_a(x) \bar{q}_b(0) \mid 0\rangle \equiv S^{ab}(x)
= \frac{i}{2\pi^2}\biggl [ \delta^{ab}
\frac{\hat{x}\cdot}{x^4}(1-AS) -BS\cdot \delta^{ab} -
\frac{x^{\alpha}\gamma^{\beta}\gamma^5}{4x^2}
\hat{G}^{ab}_{\alpha\beta}\biggr ]\ee

where
\be
AS = \frac{m_s a\delta}{96} x^4 +\frac{m_s m^2_0 a\delta}{96\cdot
24}x^6\ee

\be
BS =\frac{i}{2}{m_s}{x^2} - i \frac{a\delta}{24} \biggl (
1+\frac{m^2_0 x^2}{16} +\frac{b\cdot x^4}{4\cdot 96\cdot 3}\biggr
)\ee

\be
\hat{G}^{ab}_{\alpha\beta}=\varepsilon^{\alpha_1\beta_1\alpha\beta}
{G}^{n}_{\alpha_1\beta_1}t^n_{ab}/2\ee

and
$$ a = -(2 \pi)^2 \langle 0 \vert \bar{q} q \vert 0 \rangle$$
 $$b = \langle 0 \vert g^2 G^2 \vert 0\rangle$$
 \be
 g \langle 0 \vert \bar{q} \sigma_{\mu \nu}
 (\lambda^n/2) G^n_{\mu \nu} q \vert 0 \rangle  \equiv m^2_0 \langle
 0 \vert \bar{q} q \vert 0 \rangle
 \ee

Here $m_q=m_s$, $\delta= \langle 0 \vert \bar{s}s \vert 0 \rangle
/\langle 0 \vert \bar{q} q \vert 0 \rangle $   for s-quark case.
(It is clear, that for $u,d$ quark propagators one should put
$m_q=0$ , and unity instead $\delta$ )

We should note, that for quark propagator in the external gluon
field (the last term in (7)) we neglect the corrections,
proportional to quark mass (i.e. the term, proportional to $m_s$),
because the contribution of the terms, connected with this
corrections in the any order of OPE are smaller, than total
uncertainty (due to factorization) of the  main terms of the same
dimension. Basing on the same reason we neglect all the terms,
proportional to $m_sb$ and also some other term of high
dimensions, the contribution of which was found to be smaller than
$15\%$ of the main terms of the same dimension. (Some of such
omitted terms will be discussed below). We neglect also all the
terms, proportional to $\alpha_s$. That's why in the quark wave
function expansion (we restrict ourselves to the first five terms
of expansion)

$$ \psi (x) =\psi(0) + x^{\alpha}\nabla^{\alpha}\psi
+\frac{x^{\alpha} x^{\beta}}{2} \nabla^{\alpha}\nabla^{\beta} \psi
+\frac{x^{\alpha} x^{\beta}
x^{\gamma}}{6}\nabla^{\alpha}\nabla^{\beta} \nabla^{\gamma} \psi +
$$

\be + \frac{x^{\alpha}x^{\beta}x^{\gamma}x^{\tau}}{24} \cdot
\nabla^{\alpha}\nabla^{\beta}\nabla^{\gamma} \nabla^{\tau}\psi +
... \ee

we don't account the fourth term (with three derivatives), because
it contribution to S.R. is proportional to $\alpha_s$ due to
equation of motion, see $\cite{17}$. All terms in quark
propagator, corresponding to quark wave function expansion (11),
are accounted in BS and AS. In the simple case $m_s=0$ the terms
in the brackets in BS correspond to first, third and fifth terms
in (11).

Prior to we write the sum rules, let us discuss some common
properties of multiquark sum rules. For a moment, let us for
simplicity neglect the strange quark mass. Then it is clear from
eq.(7), that for contribution of the quark condensate of the
$n-th$ power, which have the dimension $d=3n$, there are the
contributions of dimension $d+2$ and $d+4$ (due to second and
third term in $BS$) which have fixed additional factor in
comparison with the dimension $d$ contribution. For example, if
the contribution of $d=6$ (only the part proportional to $a^2$) is
denoted as $A_6$


\be
A_6 = x^{\alpha}C/x^{10} \ee

where C is some numerical coefficient, then  $d=8$ and $d=10$
contribution from $BS$ can  be written as

\be
A_8 =(2m_0^2/16) x^{\alpha}C/x^8 \ee

\be
A_{10 }=( \frac{m_0^4}{256} + \frac{2b}{2^73^2} ) x^{\alpha}C/x^6
\ee

for any current. After Fourier and Borel transformation and
cancellation of continuum contribution with the corresponding QCD
part one can find the following relations for these contributions
\be
 A_8 = -2m^2_0 (\overline{E}_1 / \overline{E}_2)A_6\ee

and
\be
 A_{10} = ( \frac{3m_0^4}{4} + \frac{b}{3} )
 (\overline{E}_0 / \overline{E}_2)A_6\ee

where $\overline{E}_n =(M^2)^nE_n$, $M^2$ is the Borel mass square
and function $E_n$, appeared due to continuum extraction, is:

\be
 E_n(\frac{s_0}{M^2}) = \frac{1}{n!} \int\limits^{s_0/M^2}_{0} dz
 z^n e^{-z}
 \ee

So one can see, that strong cancellation should take place for the
dimension 6 and 8 contributions ($A_6$ and $A_8$), because the
factor $2m^2_0 (\overline{E}_1 / \overline{E}_2)$ is close to 1
(at any reasonable choice $s_0=~3-5GeV^2$ and $M^2=2-3GeV^2$).
Dimension 10 contribution also plays some role in this
cancellation, (though not so significant, because $A_{10}$ is
comparatively smaller) and should be also accounted. (Such
situation take place for any dimension less than 10). This type
cancellations really appear only for multiquark currents, for the
case of usual hadrons one can easily see (just in the same way),
that $A_6$ term is much more large than $A_8$ and $A_{10}$.

Of course, one should note, that $A_8$ and $A_{10}$ aren't the
only contribution of  the appropriate dimension. For example for
$d=8$ power corrections one should take into account the terms,
when a vacuum gluon line from one quark propagator is entering in
quark condensate of another quark line (see Fig. 1b ), but all
such terms are significantly smaller than $A_8$, we discuss
before, almost for all currents (except for special cases, when
$A_6=0$) and don't seriously influence on the cancellation.
\footnote {It should be noted, that for this term there also
appears  a cancellation of corresponding $d=10$ terms, which can
be proved just in the same way as we have done in our example.}

Just for the same reason for dimensions more than 10 we neglect
the contribution of the diagrams like Fig.1b, (where two vacuum
gluon lines are entering in two quark condensates) because their
contribution is found to be smaller, than uncertainty (due to
factorization hypothesis) of main terms of he same dimension (see
diagrams in fig.1). It is clear, that all this conclusions are
also valid for the terms, proportional to strange quark mass. It
is easy to see, that this cancellation becomes negligible at large
dimensions $d>11,12$, and at these dimensions one can safely
neglect the second and third terms in $BS$ in brackets in
comparison with first one. That's why the best way of calculation
of the 2-point correlator for 5-quark current is to extend OPE at
least to dimension 12,13 to avoid the uncertainty due to possible
cancellations with higher dimensions.

Another feature of multiquark currents is domination of the high
dimension correlators (see $\cite{8}$,$\cite{16}$), even in spite
the cancellation we discuss before. The reason is that lower
dimension operators are strongly suppressed because of continuum
cancellation (and also due to small numerical factor appearing
after Fourier transformation), which lead to the following factor
(for SR1, for example) $$\frac
{2^{5-2i}}{(i-1)!}E_{i-3}*(M^2)^{(i-3)}$$

Here $i=8-d/2$, and a small factor in front of E appears due to
Fourier transformation. One can see that for low dimension
operators ($d<6$) the function $E_n$, appearing due to continuum
cancellation, is extremely small (it totally compensate the large
value of $M^{2n}$). This reflects the fact, that for the multiquark
system the perturbative contribution (as well as low dimension
operators) mainly corresponds to the multiparticle states (the
region of quark-hadron duality), and not to the resonance region.
\footnote {We would like to note, that even at a special current
when contribution of $d$=6 due to quark condensate square is zero,
the contributions of high dimensions are found to be large
enough.}

Thus we can conclude that for multiquark current it is important
(unlike the usual hadron case):

1. to take into account high dimensions up to dimension 12,13,

2. if the OPE series is terminated by dimension less than 12, keep
only those terms for which the cancellation terms of the next
order are accounted - in other case the answer can be changed
drastically.

3. The pentaquark mass is determined from the ratio of eq.'s (6)
and (5). As was shown in Ref. $\cite{mm}$, in the case of
multiquark baryon currents the OPE series convergence is much
better in this ratio, than in (5) or (6).  (The QCD sum rules for
baryons at large $N_c$ were considered in $\cite{mm}$.) So, one
may believe, that  the values of the pentaquark masses found below
are reliable, if the formulated above hypothesis about the
representation of pentaquark field by local field operator is
valid.

 \vspace{1cm}

{\bf \large ~3. Sum rules}

\vspace{1cm}

Let us consider now  the sum rules (both chirality conserving and
violating, - SR1 and SR2 correspondingly, eq.'s (2,3)) for two
following currents with quantum number of  $\theta^+$ pentaquark

\be J_A =\varepsilon^{abc} \varepsilon^{def} \varepsilon^{gcf}
(u^{a^T} Cd^b ) (u^{d^T} C\gamma^{\mu} \gamma_5 d^e)\gamma^{\mu}
c\bar{s}_g\ee

\be
 J_B  = \{\varepsilon^{abc} [(d^a C \gamma_{\mu}
d^b) u^c - (d^a C \gamma_{\mu} u^b) d^c ] \cdot \bar{s}
\gamma_{\mu} \gamma_5 u + u\leftrightarrow d \}/\sqrt2  \ee

The reason of such choice of currents as characteristic examples
for $\theta^+$ sum rules is that the structure of this currents
corresponds to two most reliable cases: the first - to 3-quark
system( analogous to baryon) multiplied to $K$-meson-like part and
the second - two diquarks and antiquark. The large number of
similar currents(though not just the same) were obtained in a
number of papers (see, for example $\cite{11}$- $\cite{16}$). As
we have discussed above, the role of the high dimension
corrections is very significant. So we account power corrections
up to dimension 14,13 (with restrictions we discuss in the
previous section).  Then the final sum rules SR1,SR2 can be
written in the following form (we will further mark them SR1A and
SR2A for the current $j_a$ and SR2B and SR2B for the  current
$j_b$):

$$ M^2K_1^{(A,B)} \biggl [ \overline{E}_5 +R^{(A,B)}_4
\overline{E}_3 + R^{(A,B)}_6 \overline{E}_2 +R^{(A,B)}_8
\overline{E}_1 +R^{(A,B)}_{10} \overline{E}_0 + $$
\be
+ R^{(A,B)}_{12}/M^2 +R^{(A,B)}_{14}/M^4\biggr ] =\bar{\lambda}^2
e^{-m_{\theta}^2/M^2}\ee

$$ M^2K_2^{(A,B)} \biggl [ R^{(A,B)}_1 \overline{E}_5 +
R^{(A,B)}_3 \overline{E}_4 +R^{(A,B)}_5 \overline{E}_3
+R^{(A,B)}_7 \overline{E}_2 +R_9^{(A,B)} \overline{E}_1 +$$
\be
+ R^{(A,B)}_{11}\overline{E}_0  +R^{(A,B)}_{13}/M^2\biggr ]
=m_{\theta}\bar{\lambda}^2 e^{-m_{\theta}^2/M^2}\ee

where

$$K^A_1=\frac{32}{105}$$

$$R^A_4=\biggl (\frac{21}{16}b -7m_sa\delta\biggr )$$

$$R^A_6 =\biggl (-a^2 + \frac{m_sa m^2_0 \delta}{6}\biggr ) 35$$

$$ R^A_8 =\frac{1925}{24} m^2_0 a^2$$

$$ R^A_{10} \biggl (70m_s a^3\delta -\frac{875}{48} a^2 b
-\frac{1085}{32} m^4_0 a^2\biggr )$$

$$R^A_{12}=\biggl ( \frac{280}{9}a^4 - 105 m_sm^2_0 a^3 \delta
+\frac{2625}{192} m^2_0 a^2 b\biggr )$$

$$R^A_{14}=\biggl ( \frac{455}{18} m_s m^4_0 a^3\delta -
\frac{1085}{36} m^2_0 a^4 -\frac{105}{128} m^4_0 a^2 b\biggr )$$

$$ K^A_2 =\frac{16}{15}~~~~R^A_1 =- m_s$$

$$ R^A_3 =-2a\delta$$

$$ R^A_5 =\frac{45}{16} m^2_0 a\delta$$

$$ R^A_7 =40 m_s a^2 +\frac{25}{36} ab\delta$$

$$R^A_9 = 40 a^3 \delta - 50 m_sm^2_0 a^2 -\frac{15}{8} m^2_0
ab\delta$$

$$ R^A_{11} =- \frac{325}{6} m^2_0 a^3 \delta +10 m^4_0 m_s a^2$$
\be
R^a_{13} =\biggl ( \frac{185}{36} a^3 b\delta +\frac{145}{12}
m^4_0 a^3\delta -\frac{160}{9} m_s a^4\biggr ) \ee

$$ K^B_1 =\frac{12}{35}$$

$$K^B_2 = \frac{18}{5}$$

$$ R^B_4 =\frac{35}{18} \biggl (\frac{3}{4} b -\frac{9}{5} m_s
a(1+2\delta)\biggr )$$

$$ R^B_6 =\frac{35}{18}  ( 3 m^2_0 m_s a\delta -3a^2(1+\delta))$$

$$ R^B_8 =\frac{35}{18} a^2m^2_0 (14+3\delta) $$

$$ R^B_{10} =\frac{35}{18} \biggl ( -6 m_s a^3 (2-\delta)
-\frac{33}{4} m^4_0 a^2 + a^2 b (2+\delta/2)\biggr ) $$

$$ R^B_{12}=\frac{35}{18} \biggl ( -8 (1+\delta)a^4 + m_s m^2_0 a^3 \biggl (11 -
\frac{27}{2} \delta\Biggr ) - m^2_0 a^2 b \biggl ( \frac{11}{6}
+\frac{5}{4} \delta\biggr )$$

$$ R^B_{14} =\frac{35}{18} \biggl (m^2_0 a^4 \biggl ( \frac{22}{9}
+\frac{14}{3} \delta \biggr ) + m^4_0 m_s a^3 \biggl
(-1+\frac{97}{24} \delta \biggr ) + m^4_0 a^2 b (3/8 +3/16 \delta
) \biggr )$$

$$ R^B_1 =\frac{m_s}{6}$$

$$ R^B_{3} =\frac{1+\delta}{3} a$$

$$ R^B_5 =\frac{5}{18} (1-\delta) m^2_0 a$$

$$ R^B_7 =\frac{5}{27} \biggl ( 12 (1-\delta/2)m_s a+ (\frac{13}{2} +3\delta) b\biggr ) a$$

$$ R^B_9 =\frac{5}{27} \biggl ( 12 (1+\delta)a^2+
(\frac{-69}{2} +\frac{9}{4}\delta) m_sm_0^2a-
(\frac{9}{2} +\frac{15}{8}\delta) m_0^2b\biggr ) a$$

$$ R^B_{11} =\frac{5}{27} \biggl ( -2 (6+17\delta)m_0^2a^2+
(\frac{51}{4} +\frac{1}{4}\delta) m_sm_0^4a\biggr ) a$$

\be
R^B_{13} =\frac{5}{27} \biggl ( -4 (8+\delta)m_sa^3+
(\frac{3}{2} +\frac{25}{2}\delta) m_0^4a^2+ 2a^2b\biggr ) a\ee

Here the indexes A,B denote the sum rules for the currents $j_a$,
$j_b$ (see eq. (19,20)), $R_n$ means the contribution of the
operators of dimension $n$, and $\bar{\lambda}^2
 = (4 \pi)^8 \lambda^2$, where $\lambda$ is defined in (2).

Continuum contributions are accounted in the left-hand sides
(l.h.s) of the s.r. resulting in appearance  of the factors $E_n$.

We can easily see that the strong cancellation we discuss in the
previous section, really appears (for example $R_6$ and $R_8$
e.t.c).

The values of $\bar{\lambda}^2$, determined from eqs.(21) and (22)
with the continuum threshold  chosen as $s_0=3.5 GeV^2$ are
plotted in Fig.2a for the value of $m_{\theta}=1.54 GeV$, and the
value of $m_{\theta}$ obtained as a ratio of (22) to (21) is
plotted in Fig.2b. The parameters were taken in accord with the
recent determination of QCD condensates $\cite{18}$, $\cite{19}$
at normalization point $\mu^2=2 GeV^2$: $a=0.63 GeV^3$, $b=0.24
GeV^4$, $m^2_0=0.8 GeV^2$, $m_s=0.15 GeV$, $\delta=0.8$.

We vary the value of the continuum threshold within the range
3-4.5$GeV^2$. One can easily see, that sum rules dependence on the
Borel mass are rather stable (both SR1 and SR2) and also the value
of the pentaquark mass is also stable enough. On the other side,
the result strongly depends on the continuum threshold value. This
fact is not unexpected, such strong dependence on the continuum
threshold reflects the fact we have discussed before: in the
pentaquark case the physical threshold of the first multiparticle
state (nucleon plus kaon) in physical representation in physical
representation of 2-point correlator, (which usually is much
larger then first resonance and is included in the continuum), is
even a little smaller than the mass of the first resonance
(pentaquark $\theta^+$) mass. That's why in the pentaquark case
the "continuum threshold" is just the free parameter, which can
not be directly associated with higher resonance masses or
physical multiparticle states threshold, and it will be fixed if
one demands that the ratio of the 2-point sum rules $SR2/SR1$
should be equal to the experimental value. That's why, it seems,
that  it is principally impossible to predict the pentaquark mass
(unlike the case of usual hadrons)from 2-point sum rule for the
pentaquark case. The only thing we can find from 2-point sum rule
is: using the experimentally known value of mass, choose
appropriate currents, which give reasonable sum rules (SR1 and
SR2) and fix the such value of the parameter $s_0$  that the
$SR2/SR1$ ratio becomes about pentaquark mass -if it is possible
to do. (Of course, as soon as this parameter is fixed, it will be
used in any other sum rule with this current, according to the
usual logic of the sum rule approach). So we see, that for our
case we can conclude that both currents $j_a$ $j_b$ can be treated
as candidates to pentaquark current, because they give sum rule
with a good stability and agree with experimental value of the
pentaquark mass at $s_0$ about 3.5$GeV^2$, but one can't say, that
sum rules (21,22) themselves predict the pentaquark existence.

And yet, surprisingly, it was found to be possible to get more
information from 2-point correlator and get some predictions about
pentaquark. We will discuss this in the next section.

\vspace{1cm}

{\bf \large ~4. The estimation to the $KN$ contribution to the sum rules}

\vspace{1cm}

The main question we try to answer in this section is: is it
possible to explain the experimental results as some coupled
baryon-meson system without any one-particle intermediate state.
Let us suppose now, that  the size of the $NK$ system (nucleon and
kaon), is not too large (less than 0.5fm), so it can be
extrapolated by some interpolating current (just the same as for
pentaquark). To account the fact that this $NK$ system is not
local let us define

\be
\langle 0\mid \eta \mid nk \rangle =(\alpha \hat{P} +\beta m)
\gamma_5 v_n \ee

where  $\alpha$, $\beta$ are two unknown formfactors, $P$ is the
four-moment, carried by nuclon-kaon system, and $v_n$ is nucleon
spinor and $m$ is nucleon mass. It is clear, that if we will be
interested in the region of the order of the pentaquark mass
1.54GeV, which is close to $(m_n+m_k)^2$, (where $m_n$ and $m_k$
are nucleon and kaon masses correspondingly),  one can suppose
that formfactors are more or less constant in this region.

Then, substituting  (25) into the relation $Im \Pi =\sum \langle
0\mid\eta \mid nk \rangle \langle nk \mid \bar{\eta} \mid 0
\rangle + cont $

and integrating of the all intermediate states momenta, one can
easily obtain for physical part of sum rules

\be Im\Pi =\hat{P} \rho_1(u) +m\rho_2 (u) +
\mbox{cont.contr}\ee

where \be u =P^2,~\rho_1=\frac{c}{8\pi}\alpha^2\biggl (
\frac{u}{2} - m^2 \biggl (1-\frac{(1-z)^2}{2}\biggr ) -
\frac{m^2_k}{2} +\frac{z^2 m^2 (m^2-m^2_k)}{2u}\biggr )\ee,~

\be\rho_2=\frac{c}{8\pi}\alpha^2((z-1)m(u-zm^2)-zm\cdot m^2_k)\ee,

and $z=\beta/\alpha$,
$c=\sqrt{(1-(m_n^2+m_k^2)/u)^2-(2m_nm_k/u)^2}$

Then after Borel transformation one can easily write  sum
rules for both kinematical structures in the form, analogous to
(21,22), where all notations of the left QCD side were  defined in
(23,24).

$$ K_1^{(A,B)} \biggl [ \overline{E}_5 +R^{(A,B)}_4 \overline{E}_3
+ R^{(A,B)}_6 \overline{E}_2 +R^{(A,B)}_8 \overline{E}_1
+R^{(A,B)}_{10} \overline{E}_0 + $$
\be
+ R^{(A,B)}_{12}/M^2 +R^{(A,B)}_{14}/M^4\biggr ] \frac
{M^2}{(4\pi)^8}= \int\limits^{s_0}_{(m_n+m_k)^2} due^{-u/M^2}
\rho_1/\pi\ee

$$ K_2^{(A,B)} \biggl [ R^{(A,B)}_1 \overline{E}_5 + R^{(A,B)}_3
\overline{E}_4 +R^{(A,B)}_5 \overline{E}_3 +R^{(A,B)}_7
\overline{E}_2 +R_9^{(A,B)} \overline{E}_1 +$$
\be
+ R^{(A,B)}_{11}\overline{E}_0  +R^{(A,B)}_{13}/M^2\biggr ]\frac
{M^2}{(4\pi)^8} =\int\limits^{s_0}_{(m_n+m_k)^2} due^{-u/M^2}
\rho_2/\pi\ee

On the other side at the pure $SU3_f$ chiral limit ( when the mass
of kaon is zero) and the momentum of kaon $q=0$ the matrix element
(25) should tend to zero. It can be easily shown using the
equation of motion, that then for formfactors we have the relation
$\alpha=\beta$. Because in our case we are interesting at physical
region about pentaquark mass, which is close to $NK$ system
generation threshold $(m_n+m_k)^2$, then all momentum of particles
are rather small and it is reasonable to demand that formfactors
can differ at the same order as $SU3_f$ violation, i.e. about
$30\%$.

So for each current we obtain two different sum rules, and the
question is:  is it possible to find such values of formfactors
(more or less of the same value, i.e $z$ vary at 0.7 to 1.3, and
$\alpha$ is free), which satisfy  both sum rules (29,30). (One
should note, that the parameter $s_0$ in this case is not the same
as for the sum rules for the pentaquark case; we will vary it,
but, of course, not very far from the pentaquark mass square
region). The numerical analysis shows, that for both currents
(19,20) it found to be impossible to satisfy sum rules (29) and
(30) at the same time for all reliable values of $z$. For example
in Fig.3a,b, the value $\xi$,

\be
\xi= 1/(4^4\pi^3\alpha)^2\ee,

obtained from SR1 (29) and SR2 (30)are shown for both choices of
current $j^A$ (Fig.3a), and $j^B$ (Fig.3b), for the case $z=1.3$.
One can see, that results of SR1 and SR2 totally contradict each
other. For the smaller values of $z$ the contradiction becomes
even more sronger (for example at $z=1$ the results of sum rules
(29,30) strongly differ by modulo and moreover, have the opposite
sign). So we can conclude, that the sum rules for the currents we
consider indicate that to explain experimental results it is
necessary to have some one-particle lower state (and pentaquark is
a reliable candidate to this state) while the two-particle coupled
lower state ($NK$ system) should be excluded .

\vspace{1cm}

{\bf \large ~4. Summary}

\vspace{1cm}

The main results of this paper are the following:

 1.The features of the sum rules for 2-point correlator with 5-quark
currents are studied. It is shown, that on one side, the
contributions of high dimension operators are significant, and, on
the other side, the strong cancellation of the neighboring
dimensions contribution should take place for any choice of the
interpolating current. These features should be taken into account
to obtain sum rules correctly, as is discussed in sect.2. The best
way, from our point of view, is to take into account OPE series up
to dimension 13,14, where one can neglect the cancellations.

 2. Analysis of the sum rules for 2-point correlator with
5-quark currents for two different currents show, that it is
impossible to explain the experimental result on pentaquark,
supposing the existence only 2-particle lower state ($NK$ -system)
in the physical representation of sum rules. For this reason one
should suppose the existence of any one-particle lower state (like
pentaquark) in this mass region. Obtained sum rules (21,22) have
good stability and correspond to pentaquark mass $m_\theta$=1.54
$GeV$ at the appropriate choice of continuum threshold.

It is very necessary to note, that all conclusions and results are
obtained without a possible instanton contribution. On the other
side, it is well known, that the instanton contribution can be
significant, especially at large dimension of OPE. (see $\cite
{20}$ -  $\cite {22}$)

That's why the instanton contribution can influence the obtained
results and conclusions and we are planning to investigate this
problem in future.

Author thanks B.L. Ioffe for many significant advises and useful
discussions.

 This work is supported in part by US Civilian Research and
 Development Foundation (CRDF) Cooperative Grant Program, Project
 RUP2-2621-MO-04, RFBR grant 03-02-16209.

\begin{figure}
\epsfxsize=10cm \epsfbox{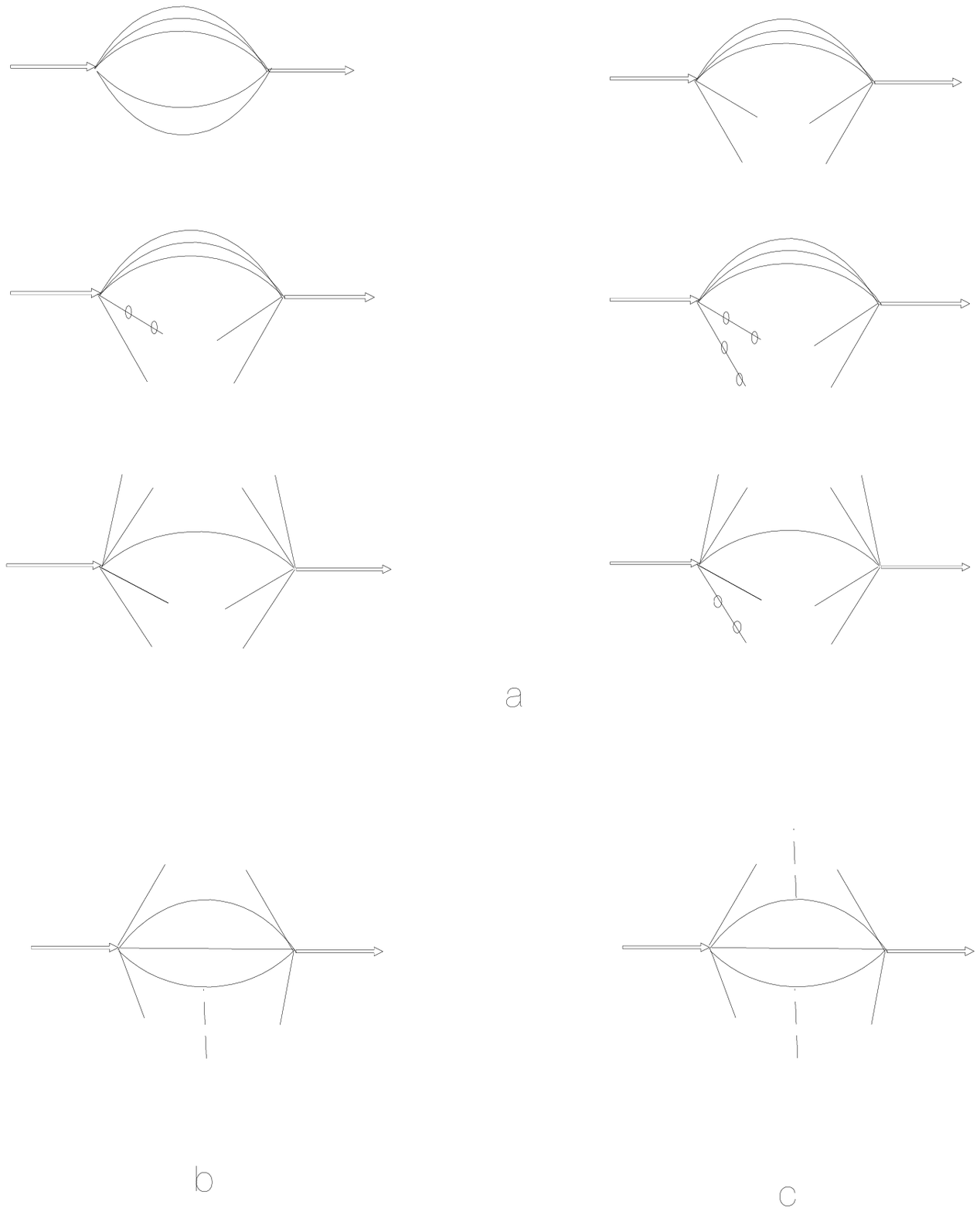} \caption{Examples of the
diagrams, corresponding to the different operators contribution.
Circles mean derivatives}
\end{figure}

\begin{figure}
\epsfxsize=14cm \epsfbox{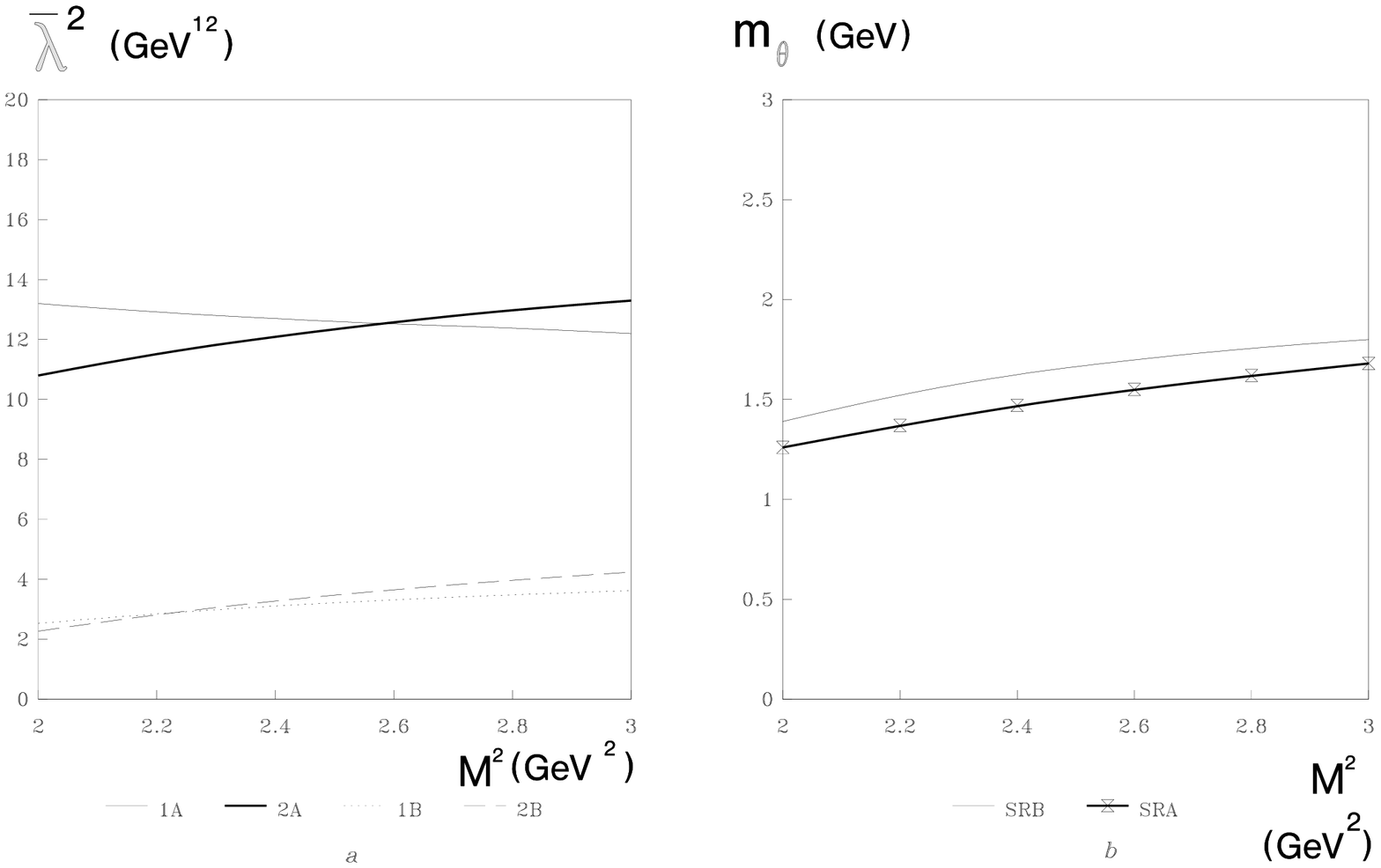} \caption{ a)Sum rules
(eqs.21,22) for currents  $j^A$ (1A and 2A correspondingly) and
$j^b$ (1B,2B). b) the pentaquark mass obtained as ratio
eq.22/eq.21 (marked SRA)for currents  $j^A$ and the same for
currents  $j^B$}
\end{figure}

\begin{figure}
\epsfxsize=14cm \epsfbox{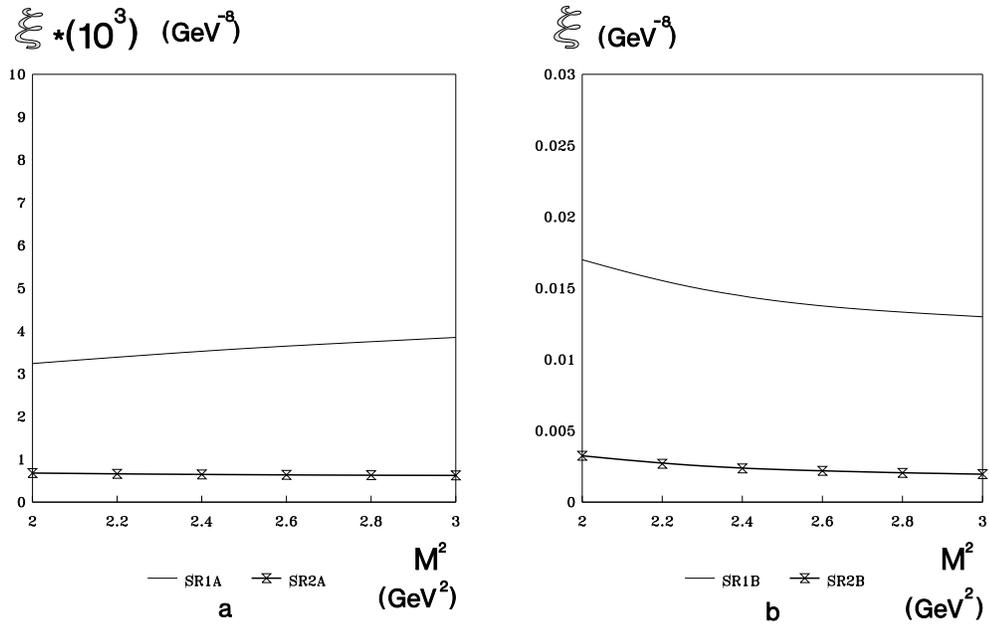} \caption{a) The value of $\xi$
(see eq.31) obtained from sum rules eq.29 (SR1A,) and eq.30 (SR2A)
for current $j^A$ . b) the same for currents $j^B$}
\end{figure}

\end{document}